\documentclass[10pt,twocolumn,twoside,submit]{JCNtran}

\usepackage[dvips]{epsfig}
\usepackage[latin1]{inputenc}
\usepackage[T1]{fontenc}

\def\BibTeX{{\rm B\kern-.05em{\sc i\kern-.025em b}\kern-.08em
    T\kern-.1667em\lower.7ex\hbox{E}\kern-.125emX}}

\begin{document}
\bibliographystyle{jcn}

\title{User Centric Content Management System for Open IPTV Over SNS (ICTC2012)}
\author{Seung Hyun Jeon, Sanghong An, Changwoo Yoon, Hyun-woo Lee, and Junkyun Choi
\thanks{Manuscript received January 31, 2013; approved for publication by Dr. Husheng Li Editor, Journal of Communications and Networks, February 13, 2015. An earlier version of this paper was awarded as best paper at the IEEE International Conference on ICT Convergence (ICTC), Jeju, Korea, October 2012.}
\thanks{This research was supported by 'The Cross-Ministry Giga KOREA Project' of The Ministry of Science, ICT and 
Future Planning, Korea [GK14P0100, Development of Tele-Experience Service SW Platform based on Giga Media] and partly supported by the ICT R\&D program of MSIP/IITP, Republic of Korea. [1391104001, Research on Communication Technology using Bio-inspired Algorithm].}
\thanks{S. H. Jeon, S. An, and J. Choi are with the Korea Advanced Institute of Science and Technology, 193 Munji-ro Yuseong-gu, Daejeon 305-732, Korea, email: \{creemur, ancom21c\}@kaist.ac.kr, jkchoi59@kaist.edu and J. Choi is the corresponding author.}
\thanks{C. Yoon and H.-w. Lee are with the Electronics and Telecommunications Research Institute, South Korea, email: \{cwyoon, hwlee\}@etri.re.kr.}}
\markboth{JOURNAL OF COMMUNICATIONS AND NETWORKS, VOL. X, NO. X, XX 2015}{Jeon \lowercase{\textit{et al}}.: User Centric Content Management System for Open ... (ICTC2012)}\maketitle

\begin{abstract}
Coupled schemes between service-oriented architecture (SOA) and Web 2.0 have recently been researched. Web-based content providers and telecommunications company (Telecom) based Internet protocol television (IPTV) providers have struggled against each other to accommodate more three-screen service subscribers. Since the advent of Web 2.0, more abundant reproduced content can be circulated. However, because according to increasing device's resolution and content formats IPTV providers transcode content in advance, network bandwidth, storage and operation costs for content management systems (CMSs) are wasted. In this paper, we present a user centric CMS for open IPTV, which integrates SOA and Web 2.0. Considering content popularity based on a Zipf-like distribution to solve these problems, we analyze the performance between the user centric CMS and the conventional Web syndication system for normalized costs. Based on the user centric CMS, we implement a social Web TV with device-aware function, which can aggregate, transcode, and deploy content over social networking service (SNS) independently. 
\end{abstract}


\section{\uppercase{Introduction}}
\label{sec:introd}
\PARstart{M}{any} studies have been conducted on concepts and relationships between service-oriented architecture (SOA) and Web 2.0 \cite {Christoph,Tim,Dirk}. SOA supports service abstraction and exposure through an enterprise service bus (ESB) and standard interfaces. Telecommunications companies (Telecoms) may make it possible to set up loosely coupled business transactions based on SOA. Web 2.0 reproduces content and information on the browser using feed technologies such as really simple syndication (RSS) and Atom Syndication Format (ATOM). In particular, Internet protocol television (IPTV) providers have aggressively adopted Web 2.0 based on service delivery platform (SDP) core technologies \cite {SDP}. Hence, telecoms can provide open application programming interfaces (APIs) related to broadcasting and communications as well as content management system (CMS) to their IPTV users or the third party providers.

Telecoms have been an efforts to provide convergence services on IPTV efficiently, compared with content providers and portal providers. However, due to increased small-sized devices (e.g., smartphone and small laptop) to watch TV content, IPTV providers have been required to support IPTV services for various mobile devices including TV. 

Conventional IPTV CMS provides reproduced content in a provider centric approach. These approaches for IPTV providers and users can rise the following problems: IPTV providers should prepare content with various file sizes regardless of whether IPTV users actually download the content. In this manner, IPTV providers cause the unnecessary depletion of CMS's storage and network bandwidth for three-screen service \cite {Zhu}, which has been provided by a prototype system of content personalization and adaptation using a PC, TV, and a mobile terminal. On the contrary, IPTV users may be forced to download oversized content by transcoding \cite {Chandra}, which leads to wasting device's storage.

In this paper, we present the method of a user centric CMS for an IPTV provider on social networking service (SNS). The main contributions of this paper are follows:
\begin{itemize}
\renewcommand{\labelitemi}{$\bullet$}
\item We propose a user centric CMS integrating SOA and Web 2.0. The following main functions of the conventional CMS are completely separated into Web services: content aggregation, mediation, and deployment. Hence, user centric CMS provides useful CMS APIs for broadcasting and communications to IPTV providers or the third party providers.

\item We analyze normalized costs for three-screen services between the proposed user centric CMS and the conventional CMS. The proposed user centric CMS for three-screen service (e.g., PC, iPad, and iPhone) is much more cost-effective than the conventional Web-based CMS because IPTV providers can reduce the depletion of CMS's processing costs and storage as well as network bandwidth for download of oversized content.

\item We develop a social Web TV to support three-screen service for open IPTV using the proposed user centric CMS. In particular, social Web TV can automatically provide content adaptive to user's device based on a device-aware function of the proposed social Web TV and content mediation of user centric CMS.
\end{itemize}

The remainder of this paper is organized as follows: we discuss related work in Section II. We propose a user centric CMS, including the system architecture, social Web TV, and three-screen service procedures in Section III. Section IV presents the system model for numerical analysis. In Section V and VI, we briefly present performance evaluations and implementation of social Web TV based on user centric CMS. Finally, we conclude the paper with directions for future work in Section VII.

\vspace{10pt}
\section{\uppercase{Related Work}}
\label{sec:related}

CMSs to support various multimedia devices have been researched for a long time. 

In social TV, Martin et al. \cite {Martin} presented a new video delivery system in social environment by integrating multiple devices such as a TV, PC, and smartphone. With their approach, after an IPTV user's request for content adaptation, the process of content reproduction starts. However, they pre-encode original content adaptive to supporting three-screen service before content aggregation. Wu et al. \cite {Wu} proposed cloud-based mobile social TV (CloudMoV). The proposed system also supports cross platform and a tanscoder of CloudMov dynamically decides encoding format in a real-time manner as well as provides proper APIs. However, CloudMov as surrogate does not manage reproduced content as CMS and contributes to reducing battery consumption for mobile users.

In content adaptation, scalable video coding (SVC) can be considered as one of technologies for the user centric CMS. Liu et al. \cite {Liu} suggested the comparison between transcoding and SVC. Though computational complexity, SVC can reduce bit-rate more than transcoding. However, SVC has limited bit-rate reduction at the base layer. Moreover, SVC is not suitable for mobile social TV \cite {Wu}. Therefore, transcoding is a good solution to be suitable for Web based CMS in converged wired and wireless access network. Furthermore, transcoding is able to fit the characteristics of the mobile device.

In network bandwidth, cost, and server overhead, Chandra et al. \cite {Chandra} mentioned that the large size of multimedia on the Internet leads to slow, expensive, and congested network. Specially, due to mobile users who access to heavy content, CMS undergoes the overhead such as time and cost. Though transcoding of heavy images, the ability of a Web service using quality aware transcoding can reduce the network bandwidth depletion as well as dynamically provide differentiated services. Atenas et al. \cite {Atenas} provided an IPTV transcoding solution to reduce the network congestion and maximize quality of experience (QoE) to IPTV users. They used a video lan client (VLC) media player as soft transcoder.

In CMS, Yang et al. \cite {Jinhong} presented a Web-based content syndication platform for IPTV providers or the third party providers, which will be called as content aggregation and syndication system (CANSS). As Web application, CANSS provides brief pre-encoded content to IPTV providers or the third party providers before contracts to sell transcoded content. Finally, after purchasing the transcoded content to be reproduced and deciding transforming method, they receive transcoded content from CANSS. Hence, their systems supply reproduced content in a pre-encoding manner. Moreover, CANSS aggregate content from feeds of content providers, convert aggregated content, and deploy content in a sequential and automatic manner. Li et al. \cite {Li} proposed Cloud Transcoder, which bridges the format and resolution gap between Internet videos and mobile devices. After a mobile user uploads a video request including video link, format, and resolution to Cloud Transcoder content aggregation, transcoding, and deployment are sequentially worked to transfer transcoded content. Though the real-time transcoding, the repeated transcoding requests for popular content are still burden for Cloud Transcoder.

However, we separate the integrated syndication system into independent Web services and can request them on social environment in a real-time manner. 

\vspace{10pt}
\begin{figure}[t]
\begin{center}
\epsfxsize=8.8cm \leavevmode\epsfbox{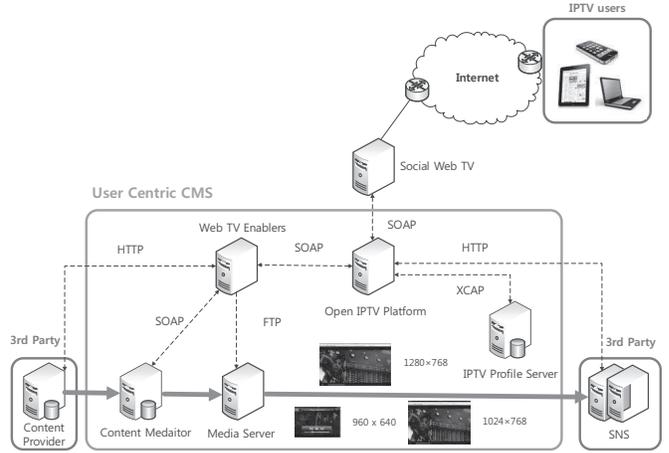}
\caption{User centric CMS with social Web TV.} \label{fig:system model}
\end{center}
\end{figure}
\begin{figure}[t]
\begin{center}
\epsfxsize=8.8cm \leavevmode\epsfbox{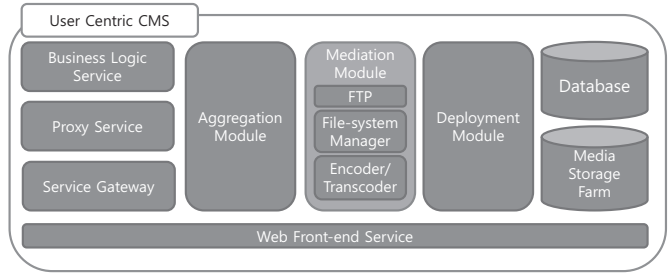}
\caption{Logical block diagram of user centric CMS.} \label{fig:block diagram}
\end{center}
\end{figure}
\section{\uppercase{Proposed User Centric CMS}}
\label{sec:proposed}
\subsection{System architecture}
First, we newly design content aggregation, transcoding, and deployment suitable for open IPTV. The tightly-coupled Web syndication platform is disjointed as follows: content aggregation, mediation, and deployment. Each function is called from the Web independently, and then is open as Web service to the Internet. Furthermore, they may be easily combined with other Web 2.0 technologies such as eXtensible Markup Language (XML), RSS, and ATOM. The proposed user centric CMS consists of Web TV enablers, a content mediator, a media server, an IPTV profile server, and an open IPTV platform in Fig. \ref {fig:system model}. The main functions for user centric CMS are operated at Web TV enablers, which include content aggregation, mediation, and deployment. Simple object access protocol (SOAP) is used for delivering XML messages between Web TV enablers and an open IPTV platform in Fig. \ref {fig:system model}. The remaining interfaces are explained in Fig. \ref {fig:system model}. SOAP as one of Web services in SOA is standardized by the World Wide Web Consortium (W3C) \cite {soap}. We assume that devices of IPTV users are divided into the following three types: PC, iPhone, and iPad.

Fig. \ref{fig:block diagram} illustrates a logical block diagram of our system configuration of the user centric CMS. User centric CMS consists of several service logic modules, media storage farm, and three main function modules of aggregation, mediation, and deployment with supporting database system. Mediation module includes file transfer protocol (FTP) service for file transferring, file system manager, and encoder/transcoder.

As security concern, we allow to access to user centric CMS for IPTV users acquiring authentication from social Web TV. Anonymous social users are restricted to the user centric CMS.
\subsubsection{Web TV enablers}
Web TV enablers consist of content aggregation, content mediation, and content deployment. 

Content aggregation collects content from content providers. As a Web content feeder based on RSS and Web syndication format (ATOM) a content provider may upload content suitable to formats, and then IPTV providers or the third party providers can choose and aggregate content. Original content will temporarily be aggregated at a content mediator. FTP is required for content delivery to temporary storage.

Content mediation provides content transcoding for aggregated content. We adopt FFmpeg as a software encoder. FFmpeg provides an audio/video converting solution for a cross platform as free licensed software \cite {ffmpeg}. Besides, content mediation interworks with DeviceID in device profile of an IPTV profile server in order to acquire audio/video encoding information suitable for user's device. DeviceID is used as an identifier to classify IPTV user's device. An example of device profile is shown in Fig. \ref{fig:device profile} of Appendix. After completing content encoding, newly generated content is assigned a new content reference identifier (CRID) \cite {tv-anytime}, which is generated by the IPTV profile server. If transcoded content with the same device profile by another IPTV user already exists, additional content mediation does not work despite different device by using isExistContent Web service in Table \ref{tab:core Web services} of Appendix. CRID is structured as uniform resource locator (URL) and is described such as \textit{'crid://etri.re.kr/webtv/20120602'}. The last serial number is generated considering date. Filename of transcoded content is structured as \textit{'original content filename + CRID's last number + device profile's video encoding'} in Fig. \ref {fig:ws_transcodeContent} of Appendix.

Content deployment delivers transcoded content to a media server, where the converted content is stored. Before delivering content, connection in a FTP manner is required. If the converted content is new, metadata generated in TV-Anytime \cite {tv-anytime} should be inserted into the content profile of an IPTV profile server. Content deployment can support content sharing by delivering a URL for content location. Using uploading APIs provided by Twitter and me2day, the transcoded content can be delivered to other social services (me2day is a SNS operated by NHN in South Korea).

Web TV enablers are operated by Oracle Weblogic server. 
\subsubsection{Content mediator}
A content mediator is a transcoding engine, which employs FFmpeg open sources, and interworks with Web TV enablers. FFmpeg is installed on an Ubuntu server operating system (OS). This system temporarily stores the original content aggregated by content providers and converted content generated by content mediation.
\subsubsection{Open IPTV platform}
Open IPTV platform is a gateway to exchange XML information to harmonize APIs such as Web TV enablers and device profile of an IPTV profile server in Fig. \ref {fig:system model}. Web TV enablers can be open to IPTV providers or the third party providers through ESB. We use Oracle Communications Service Gatekeeper (OCSG) as the core SDP for open IPTV.  
\subsubsection{IPTV profile server}
An IPTV profile server contains profiles for open IPTV related to content and devices. The open IPTV platform can communicate with the IPTV profile server in a XML Configuration Access Protocol (XCAP) manner of Fig. \ref {fig:system model}. XCAP is an application protocol, which can insert, modify, and delete XMLs with a database. CRID generation and management are controlled by content profile of the IPTV profile server. Device profile of the IPTV profile server consists of DeviceID, resolution, and video/audio encoding information like Fig. \ref {fig:device profile} in Appendix. 

IPTV providers can decide the optimal resolution per each device for three-screen service. Even though a mobile device supports various resolutions, IPTV providers can choose the most optimal resolution and encoding information for three-screen service considering display specification of manufacturers, IPTV's user's preference and server operation costs.

On the contrary, if an administrator should transcode content regardless of mobile device's similarity, separate content transcoding can be possible, as adjusting detailed device profile such as audio and video encoding information.
\begin{figure}[t]
\begin{center}
\epsfxsize=8.8cm \leavevmode\epsfbox{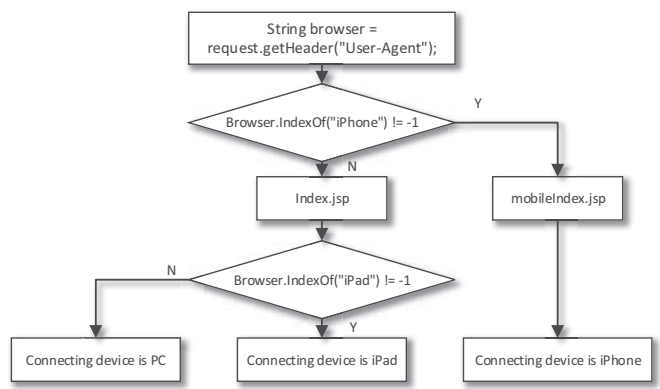}
\caption{Device-aware function.}
\label{fig:device-aware}
\end{center}
\end{figure}
\subsubsection{Media server}
A media server is a Web server for content storage as well as a FTP server for content management on Internet Information Server (IIS) provided by Microsoft.
\subsection{Social Web TV}
We also propose social Web TV using the user centric CMS for open IPTV, as seen in Fig. \ref{fig:system model}. The proposed social Web TV integrates an operation portal for an administrator with SNS for IPTV users. Requests for content aggregation, mediation, and deployment can be controlled and monitored by IPTV users over the proposed social Web TV. The traditional SNS only focuses on the content aggregation, and deployment. However, because social Web TV provides IPTV users with the ability of content mediation and device-aware function additionally, IPTV providers may accommodate more subscribers for three-screen service.

In detail, we present device-aware function and philosophy as well as pros and cons for social Web TV.  
\subsubsection{Device-aware function}
First of all, we add a device-aware function into social Web TV, which automatically detects three device types through the initial connection such as login page. In Fig. \ref{fig:device-aware}, the device-aware function can recognize the device type from the user-agent message of the hypertext transfer protocol (HTTP) header on the browser. At the device-aware function, social Web TV separately provides a mobile Web page for iPhone. PC uses the same Web page as iPad because PC and iPad can support high resolution like TV.
Therefore, content mediation of the user centric CMS using the device-aware function of social Web TV can automatically provide the optimal transcoded content without IPTV user's requests.

If the user-agent of HTTP header can support detailed device types, more accurate device-aware function can be developed. How accurate to detect user devices and how various to support user devices are dependent on embraced device information within the user-agent message. Therefore, device-aware function has an enough chance to accommodate various user devices in Fig. 3.

\subsubsection{Philosophy}
Increasing social media has high capacity for content diffusion among SNSs. Synergy between TV with social media is also expected by delivering better content precisely and rapidly. Social TV should provide good accessibility and screen diversity (three-screen), as there are various kinds of social users with different screens and access environment.

Among the approaches to this problem, World Wide Web (WWW) is particularly suited to provide accessibility. Moreover, we can apply the proposed user centric CMS to social service environment independently of Web browsers. Generally, SNS can provide aggregated content to familiar social users as a Web content aggregator. Content reproduction based on social environment has not been permitted by service providers, because content modifications are restricted by content suppliers. However, the proposed social Web TV enables to aggregate content, transcode aggregated content, and deploy the transcoded content. For example, user created content (UCC) is free of limitation for content transcoding. The restriction of content reproduction by the content suppliers is beyond the scope of this paper.

\subsubsection{Advantages}
First, the proposed social Web TV can reproduce content according to IPTV user's requests. Besides, social Web TV can support any mobile devices including a PC if plentiful device profile is stored and device-aware function is delicately extended. IPTV users can freely aggregate content from content providers and then transcode original content when they want to watch them. Second, the proposed user centric CMS can reduce the depletion of network bandwidth and storage from being exhausted by downloading high quality content. Third, content aggregation, mediation, and deployment work on the same social portal individually. This enables the open IPTV platform to provide IPTV users or the third party providers with Web TV enablers of the user centric CMS as open API. Lastly, since IPTV users do not request content mediation for unpopular content, preparation of content transcoding for three-screen service is unnecessary.

\subsubsection{Disadvantages}
At least one voluntary IPTV user needs to request content transcoding. The initial requester should wait for watching the transcoded content until the completion of content transcoding. Next, administrators for media servers should monitor file sizes, because the number of reproduced content may increase in proportion to the number of device profile. Hence, the proposed social Web TV provides IPTV users with important information relating to the maintenance of the user centric CMS such as counts of content mediation, content viewing, and the total content.
\subsection{Three-screen service procedures}
Based on social Web TV, which accommodates the proposed user centric CMS, we present several three-screen service procedures. Interfaces and operations of core Web services for the user centric CMS in Table \ref{tab:core Web services} of Appendix are shown.
\subsubsection{Procedure of content aggregation}
\begin{figure}[t]
\begin{center}
\epsfxsize=8.8cm \leavevmode\epsfbox{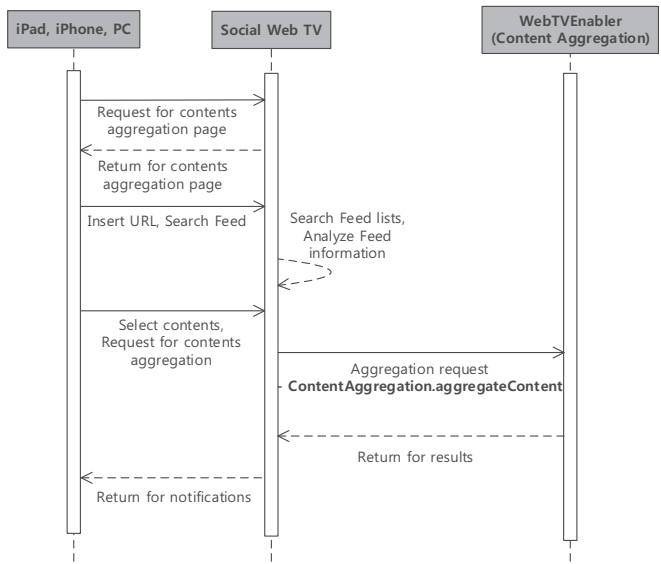}
\caption{Procedure of content aggregation on social Web TV.}
\label{fig:procedure of content aggregation}
\end{center}
\end{figure}

In this procedure, an IPTV user wants to see a set of content from a certain content feed (e.g., RSS or ATOM). The IPTV user inputs the address of the feed on the aggregation page of social Web TV, and then the Web TV enabler reads the feed to aggregate content. The IPTV user can select the content to be aggregated. The procedure follows the sequence of content aggregation described in Fig. \ref{fig:procedure of content aggregation}.

- IPTV user connects to the aggregation page of social Web TV through a PC/iPad

- IPTV user inputs the URL of the content feed of RSS or ATOM

- By searching content, Web TV enabler reads, analyzes aggregated feeds, and then returns the list of content to IPTV user

- IPTV user selects content to be aggregated, from the list of results

- Social Web TV requests aggregation to the Web TV enabler by using ContentAggregation.aggregateContent

- After finishing the aggregation, the Web TV enabler returns the aggregation process results

- Social Web TV notifies the results to IPTV user
\begin{figure}[t]
\begin{center}
\epsfxsize=8.8cm \leavevmode\epsfbox{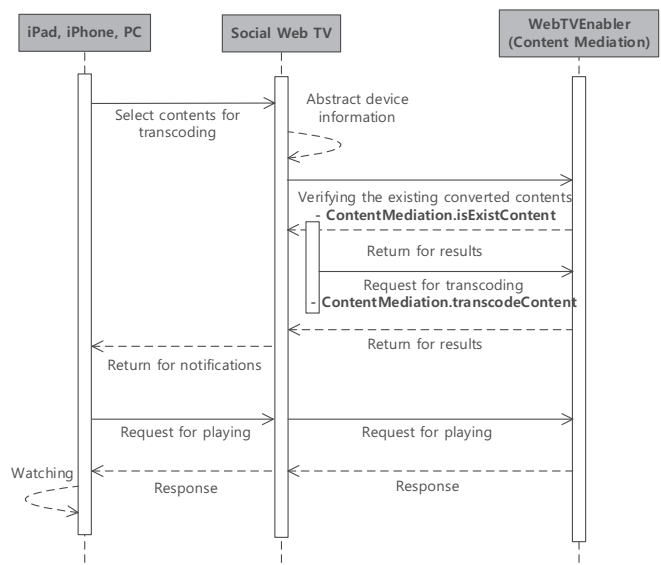} 
\caption{Procedure of content mediation on social Web TV.} 
\label{fig:procedure of content mediation}
\end{center}
\end{figure}
\begin{figure}[t]
\begin{center}
\epsfxsize=8.8cm \leavevmode\epsfbox{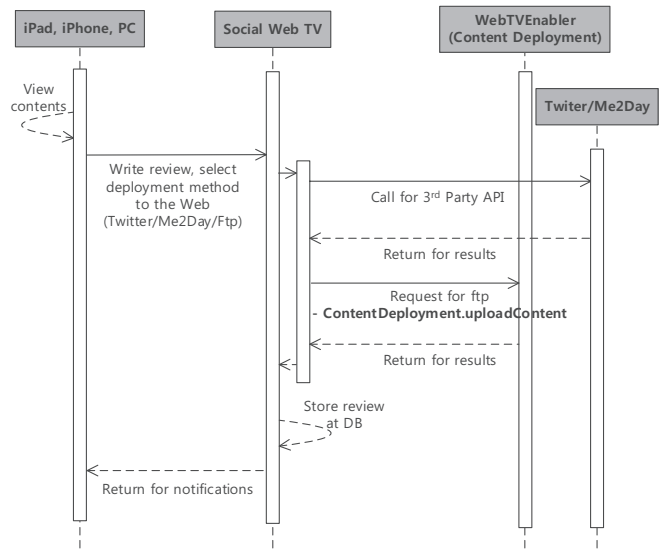}
\caption{Procedure of content deployment on social Web TV.}
\label{fig:procedure of content deployment}
\end{center}
\end{figure}
\subsubsection{Procedure of content mediation}
In this procedure, an IPTV user wants to watch the optimal content. Social Web TV checks the user's device environment through the initial connection in Fig. \ref{fig:device-aware} and then confirms whether the Web TV enabler has properly converted content. If not, social Web TV asks the IPTV user to transcode the content. Even if the IPTV user revokes content transcoding, the IPTV user can watch the original content. If the content has already been converted, the IPTV user can instantly watch the content without request message for content mediation. The procedure of content mediation follows the sequence described in Fig. \ref{fig:procedure of content mediation}.

- IPTV user selects one of the content lists through an iPhone

- Social Web TV checks IPTV user's device information and then recognizes IPTV user's device as an iPhone

- Social Web TV verifies the converted content suitable to IPTV user's device, by using ContentMediation.isExistContent

- Web TV enabler notifies that there is no converted version for an iPhone

- Social Web TV notifies to IPTV user that there is no converted version, and asks IPTV user to convert the content by using ContentMediation.transcodeContent

- If IPTV user agrees to convert the content, social Web TV requests transcoding of the content

- After finishing transcoding, social Web TV notifies that it is ready to play the content

- IPTV user plays the content

\subsubsection{Procedure of content deployment}
In this procedure, an IPTV user wants to share the watching experience of social Web TV with a SNS. The social Web TV supports social deployment with watched content information. The procedure of content deployment follows the sequence described in Fig. \ref{fig:procedure of content deployment}.

- IPTV user with an iPad connects to social Web TV

- IPTV user writes a review of content that IPTV user watched on social Web TV before

- IPTV user requests to deploy converted content to Twitter and me2day (content sharing)

- Social Web TV calls the third party APIs such as Twitter or me2day in order to post a review with content information and upload the content

- Social Web TV shares the content by using ContentDeployment.uploadContent

- Social Web TV notifies the result to IPTV user
\vspace{10pt}
\section{\uppercase{System Model}}
\label{sec:system}
We assume the popularities (i.e., access probability) of content are described by a Zipf-like distribution \cite{Wallace,Lei} as follows:
\begin{eqnarray}\label{a}
P_r  = \frac{C}{{r^\delta  }},
\end{eqnarray}
where 
\begin{eqnarray}\label{b}
C = \left( {\sum\limits_{r = 1}^N {\frac{1}{{r^\delta  }}} } \right)^{ - 1},
\end{eqnarray}
 as a constant for a Zipf-like distribution, $N$ is the number of content items. In particular, $\delta=0.271$ is applied by \cite {Wallace}. 
We assume that popular content has high access probability to provide IPTV subscribers with a three-screen service. Original content may be transcoded dependent on Table \ref{tab:three-screen service scenarios}. 
\begin{table}[t]
\caption{Three-screen service scenarios for content mediation.}
\label{tab:three-screen service scenarios}
\begin{center}
\begin{tabular}{ p{1.5cm} |  p{3cm} | p{3cm} }
    \hline
    Scenarios & Original content & Converted content\\ \hline
    $s_{1}$ & PC & iPad \\ 
    $s_{2}$ & PC & iPhone \\ 
    $s_{3}$ & iPad & iPhone \\ \hline
\end{tabular}
\end{center}
\end{table}

\begin{table}[t]
\caption{Configuration parameters.}
\label{tab:configuration parameters}
\begin{center}
\begin{tabular}{ p{1.6cm} |  p{4.3cm} | p{1.6cm} }
    \hline
    Parameters & Descriptions & Values\\ \hline
    $r$ & Video ranks & 1$\le$r$\le$1000 \\ 
    $N$ & The number of content & 1000 \\ 
    $U$ & The number of three-screen service subscribers & 10000 \\ 
    $C_{agg}$ & The normalized costs of content aggregation & 0.2 \\ 
    $C_{med}$ & The normalized costs of content mediation & 0.7 \\ 
    $C_{dep}$ & The normalized costs of content deployment & 0.1 \\ 
    $\alpha$ & The normalized server loads of content mediation from PC to iPad & 0.5 \\ 
    $\beta$ & The normalized server loads of content mediation from PC to iPhone & 0.3 \\ 
    $\gamma$ & The normalized server loads of content mediation from iPad to iPhone & 0.2 \\
        $i$ & Index number for three-screen service scenarios & $\{ 1,2,3\}$ \\ \hline
\end{tabular}
\end{center}
\end{table}

\begin{table}[t]
\caption{Performance parameters.}
\label{tab:performance parameters}
\begin{center}
\begin{tabular}{ p{1.6cm} |  p{6.4cm} }
    \hline
    Parameters & Descriptions \\ \hline
    $C^{1}_{CANSS}$ & The normalized cost of CANSS based on $s_{3}$ scenario \\
    $C^{2}_{CANSS}$ & The normalized cost of CANSS based on $s_{2},s_{3}$ scenarios \\ 
    $C^{3}_{CANSS}$ & The normalized cost of CANSS based on $s_{1},s_{2},s_{3}$ scenarios \\ 
    $C^{1}_{proposed}$ & The normalized cost of the proposed scheme based on $s_{3}$ scenario \\
    $C^{2}_{proposed}$ & The normalized cost of the proposed scheme based on $s_{2},s_{3}$ scenarios \\  
    $C^{3}_{proposed}$ & The normalized cost of the proposed scheme based on $s_{1},s_{2},s_{3}$ scenarios \\ \hline 
\end{tabular}
\end{center}
\end{table}

Using configuration parameters in Table \ref{tab:configuration parameters}, the normalized costs of the three-screen service scenarios for CANSS and the proposed user centric CMS are obtained as the following performance parameters in Table \ref{tab:performance parameters}:

\begin{itemize}
\renewcommand{\labelitemi}{$\bullet$}
\item Three-screen service scenarios for CANSS
\begin{eqnarray}\label{c}
C_{CANSS}^{1}  = C_{agg} + C_{med}  \times \gamma + C_{dep},
\end{eqnarray}
\begin{eqnarray}\label{d}
C_{CANSS}^{2}  = 2 \times ( C_{agg} + C_{med}  \times \left( {\beta + \gamma} \right) + C_{dep} ),
\end{eqnarray}
\begin{eqnarray}\label{e}
C_{CANSS}^{3}  = 3 \times ( C_{agg} + C_{med}  \times \left( {\alpha + \beta + \gamma} \right) + C_{dep} ).
\end{eqnarray}

\item Three-screen service scenarios for the proposed user centric CMS
\begin{eqnarray}\label{f}
\begin{array}{l}
C_{proposed}^{1}  = C_{agg} + C_{med} \times \gamma + C_{dep}, 
 \end{array}
\end{eqnarray}
\begin{eqnarray}\label{g}
\begin{array}{l}
C_{proposed}^{2}  = C_{agg} + C_{med} \times \left( {\beta + \gamma} \right) + C_{dep}  \times 2, 
 \end{array}
\end{eqnarray}
\begin{eqnarray}\label{h}
\begin{array}{l}
 C_{proposed}^{3}  = C_{agg} + C_{med}  \times \left( {\alpha  + \beta  + \gamma } \right) + C_{dep}  \times 3.  \\ 
 \end{array}
\end{eqnarray}
\end{itemize}

Probability density functions (PDFs) between CANSS and the proposed scheme considering IPTV subscribers and popularity for each scenario using Table \ref{tab:performance parameters} and index number $i$ for three-screen service scenarios in Table \ref{tab:configuration parameters} are explained as follows:
\begin{eqnarray}\label{i}
f(C_{CANSS}^{i} ) = C/r^\alpha   \times U \times C_{CANSS}^{i}, 
\end{eqnarray}
\begin{eqnarray}\label{j}
f(C_{proposed}^{i} ) = C/r^\alpha   \times U \times C_{proposed}^{i}.
\end{eqnarray}

Cumulative distribution functions (CDFs) between CANSS and the proposed scheme are explained using equation from (9) to (10) as follows:
\begin{eqnarray}\label{k}
F(C_{CANSS}^{i} ) = \sum\limits_{r = 1}^N {C/r^\alpha   \times U \times C_{CANSS}^{i} },
\end{eqnarray}
\begin{eqnarray}\label{l}
F(C_{proposed}^{i} ) = \sum\limits_{r = 1}^N {C/r^\alpha   \times U \times C_{proposed}^{i} }.
\end{eqnarray}
\vspace{10pt}
\section{\uppercase{Performance Evaluations}}
\label{sec:imple}
We show performance evaluations based on system model and comparison between the conventional CMS and the proposed system.
\subsection{Performance analysis}
\begin{figure}[t]
\begin{center}
\epsfxsize=8.8cm \leavevmode\epsfbox{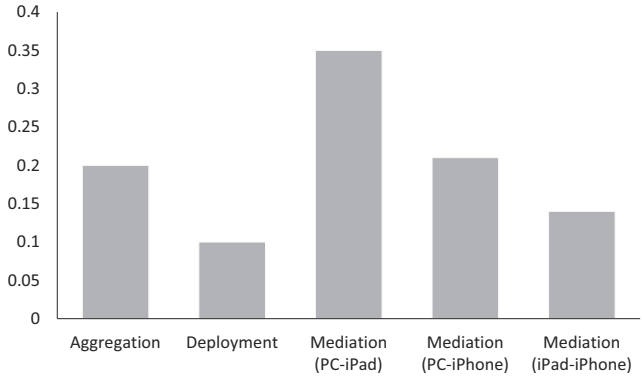}
\caption{Expected normalized costs.}
\label{fig:expected normalized costs}
\end{center}
\end{figure}
\begin{figure}[t]
\begin{center}
\epsfxsize=8.8cm \leavevmode\epsfbox{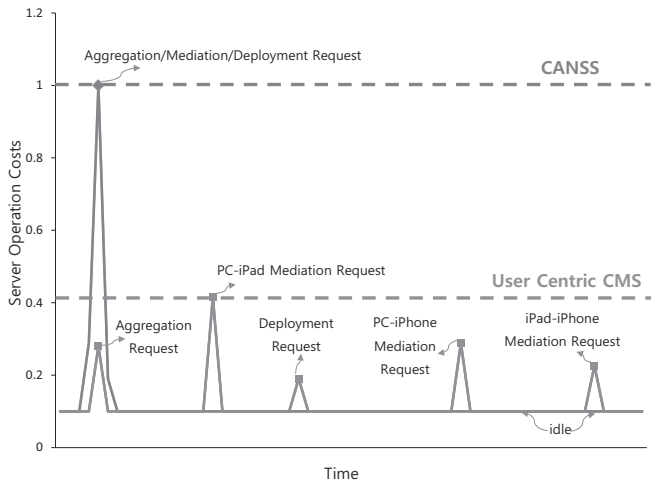}
\caption{Server operation costs.}
\label{fig:server operation costs}
\end{center}
\end{figure}
In Fig. \ref{fig:expected normalized costs}, expected normalized costs for each operation are described. We assume the sum of each operation cost is normalized as 1. The expected normalized costs relating to server operation include server overheads for resource process of CMS such as network bandwidth, storage, and transcoding time. Through repeated experiments, it is observed that the operation of content mediation causes excessive overhead in a content mediator's central processing unit (CPU), because transcoding content is a heavy task \cite {Virag}. In particular, transcoding content from a PC size into an iPad size takes much more time than other transcoding scenarios. Since content aggregation is operated at a significant distance from content provider's Web/FTP servers, the expected normalized cost for aggregation is slightly higher than that for deployment. We set the expected normalized cost for mediation (iPad-iPhone) is slightly lower than aggregation due to the same formats for $s_{3}$ scenarios such as H.264. However, all expected normalized costs are dependent on content's file sizes and compression of encoding/decoding formats.

Fig. \ref{fig:server operation costs} shows the change of server operation costs by time when content is aggregated, mediated, and deployed. The minimum operation cost (0.1) is assumed during server idle time, and it is also included for the cost of each operation in Fig. \ref{fig:expected normalized costs}. At aggregation time, CANSS takes the most server overhead, because it runs all operations of aggregation, mediation, and content deployment. However, user centric CMS independently runs the operation of content aggregation, mediation, and deployment based on the response to IPTV user's requests.
\begin{figure}[t]
\begin{center}
\epsfxsize=8.8cm \leavevmode\epsfbox{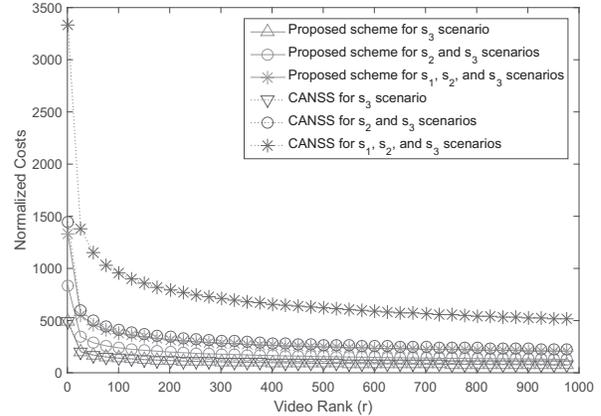}
\caption{Normalized costs.}
\label{fig:normalized costs}
\end{center}
\end{figure}
\begin{figure}[t]
\begin{center}
\epsfxsize=8.8cm \leavevmode\epsfbox{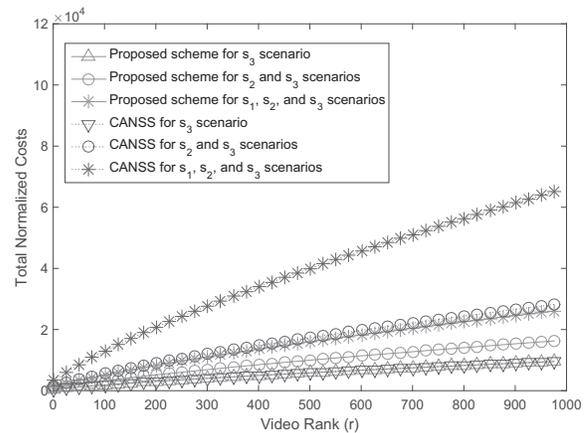}
\caption{Total normalized costs.}
\label{fig:total normalized costs}
\end{center}
\end{figure}

Through equation (9) to (10), Fig. \ref {fig:normalized costs} shows available three-screen service scenarios for the proposed scheme and CANSS. Using Table \ref{tab:three-screen service scenarios}, we consider normalized costs according to frequencies of transcoding. Fig. \ref{fig:normalized costs} shows that because popular content has higher access probability for three-screen service, normalized costs are much increased. In Fig. \ref{fig:normalized costs}, the proposed scheme achieves lower normalized costs than CANSS. However, Fig. \ref{fig:normalized costs} shows the same normalized costs for $s_{3}$ scenario due to making no difference on the frequency of content aggregation, mediation, and deployment. According to increasing transcoding frequency, the gap for normalized costs is large, because redundant requests for content aggregation in the proposed scheme are minimized. 

Fig. \ref{fig:total normalized costs} shows the total normalized costs for the proposed scheme and CANSS using equation (11) to (12). When the proposed user centric CMS provides IPTV users with three-screen service, IPTV providers can effectively reduce operation costs as well as save used network bandwidth for three-screen service.
\subsection{Comparison between the conventional CMS and the proposed system}
We compare CANSS with social Web TV based on the user centric CMS in Table \ref{tab:comparison}. The existing Web syndication platform supports a business-to-business (B2B) model for the third party provider. However, the proposed social Web TV additionally supports a customer-to-business (C2B) business model for IPTV users. The user centric CMS is completely compatible with open IPTV based on SOA. Hence, the user centric CMS can be easily adapted to an open IPTV platform. Social Web TV based on user centric CMS fully supports three-screen service as a cross platform. Through device profile, social Web TV can be recognized for IPTV user's device. For IPTV providers, waste of network bandwidth can be reduced. Content is transcoded by IPTV user's requests in a real-time manner. IPTV users can share the transcoded content with other social communities. Finally, an extra management portal for CMS is not required.
\begin{table}[t]
\caption{Comparison between CANSS and the proposed system.}
\scriptsize
\label{tab:comparison}
\begin{center}
\begin{tabular}{ p{2.3cm} |  p{2.7cm} | p{2.5cm} }
    \hline
    Items & CANSS & User Centric CMS Based Social Web TV \\ \hline
    Business Model & B2B only & B2B and C2B \\ \hline   
    SOA & Tightly-coupled & Loosely-coupled \\ \hline
    Open API & Content mediation & Content aggregation\\ & & Content mediation\\ & & Content deployment \\ & &(Content sharing) \\ \hline
    Cross Platform & Partially supported & Fully supported \\ \hline
    Device-Aware & Not supported & Supported by profiles \\ \hline
    Network Bandwidth & Slightly waste  & No waste \\ \hline
    Transcoding & Pre-encoding & Real-time encoding \\ \hline
    Reproduction's Role & Administrator & IPTV users \\ \hline
    Content Deployment & FTP & Twitter \\ & & me2day\\ & & FTP \\ \hline
    Application Server & Management portal & Social Web TV (SNS)  \\ \hline
\end{tabular}
\end{center}
\end{table}
\section{\uppercase{Implementation}}
\begin{figure}[t]
\begin{center}
\epsfxsize=4cm \leavevmode\epsfbox{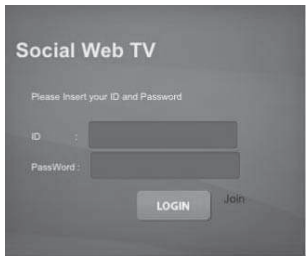}
\caption{A main page of social Web TV.}
\label{fig:login main page}
\end{center}
\end{figure}
\begin{figure}[t]
\begin{center}
\epsfxsize=6.89cm \leavevmode\epsfbox{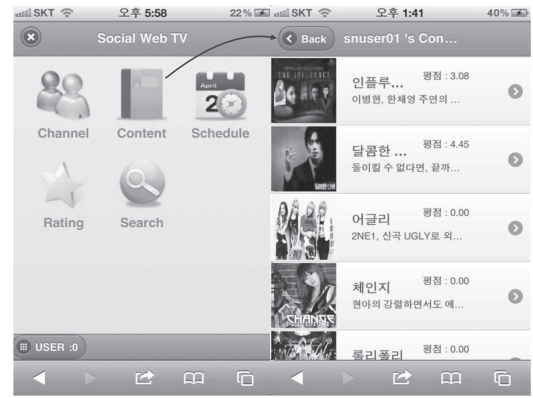}
\caption{Social Web TV for iPhone.}
\label{fig:mobile social Web TV}
\end{center}
\end{figure}
In this paper, because we focus on integrating a user centric CMS with social environment, we do not describe features or technologies for social services in detail. IPTV users aggregate, transcode, and deploy content independently on social Web TV. Social Web TV are tested on Google Chrome or Apple Safari. However, social Web TV is basically designed as cross platform using XML standard. Fig. \ref{fig:login main page} shows a main page of social Web TV based on the user centric CMS. According to user device, user pages for social Web TV are changed after login in Fig. \ref{fig:login main page}. Fig. \ref{fig:mobile social Web TV} shows social Web TV for iPhone. Each device's optimal resolution in device profile is set as the followings: 1280$\times$768 for PC, 1024$\times$768 for iPad, and 960$\times$640 for iPhone. iPhone4 and first-generation iPad are considered for iPhone and iPad, respectively. Audio/video encoding are set as freeware advanced audio coder (faac) and H.264, respectively.
\begin{figure}[t]
\begin{center}
\epsfxsize=8.8cm \leavevmode\epsfbox{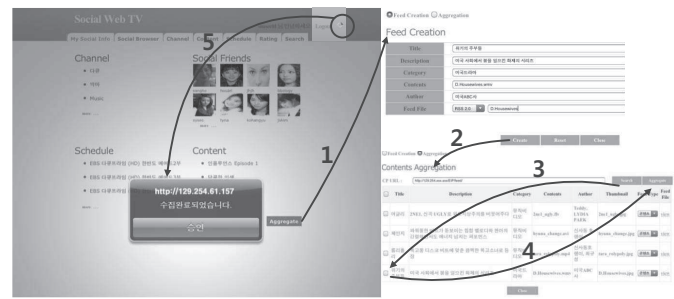}
\caption{Content aggregation in PC/iPad.}
\label{fig:aggregatecontent}
\end{center}
\end{figure}
\begin{figure}[t]
\begin{center}
\epsfxsize=8.8cm \leavevmode\epsfbox{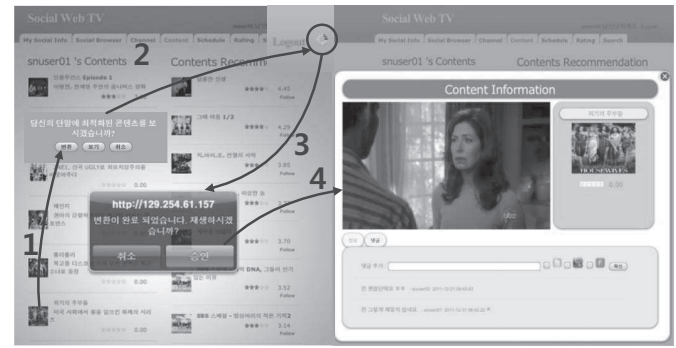}
\caption{Content mediation in PC/iPad.}
\label{fig:isexistcontent}
\end{center}
\end{figure}
\begin{figure}[t]
\begin{center}
\epsfxsize=8.8cm \leavevmode\epsfbox{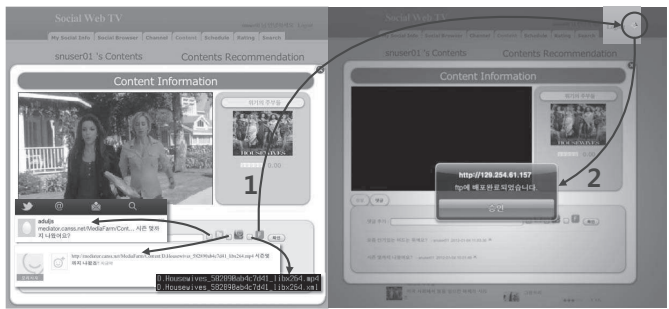}
\caption{Content deployment in PC/iPad.}
\label{fig:uploadcontent}
\end{center}
\end{figure}
In deployment concern, since a centralized mediator is assumed, bottleneck of social Web TV may occur by rushed IPTV user's requests to user centric CMS. In this case, distributed mediators can be required for user centric CMS. Also, duplicate content by database system and load balancing for IPTV user's requests should be maintained.

Fig. \ref{fig:aggregatecontent} shows content aggregation over social Web TV. First, a feed file for original content is created. Then, after searching target content for content aggregation, content aggregation is worked. Progress of content aggregation is shown on the top right of the screen. After finishing content aggregation, an IPTV user receives the response message for completion.

Fig. \ref{fig:isexistcontent} shows content mediation over social Web TV. First, after choosing content to watch, IPTV users receives the request message on whether content adaptive to user device will be transcoded or not. If transcoded content by other IPTV users already exists, this request message is not shown. After requesting content mediation, content mediation is worked. If an IPTV user watches the original content, content will be watched over the independent media player. After the completion of transcoding, an IPTV user can see transcoded content.

Fig. \ref{fig:uploadcontent} shows content deployment over social Web TV. While viewing content, an IPTV user can deploy content with other third party providers by FTP. Besides, an IPTV user can share content with social friends of Twitter or me2day. For progress of content deployment, a FTP or SNS account is required. After finishing content deployment, an IPTV user can receive the response message for completion.

Therefore, user centric CMS enables IPTV users to be familiar with SNS through social Web TV and to distribute transcoded content to social friends.
\section{\uppercase{Conclusion and Future Work}}
\label{sec:conclusion}
We proposed a user centric CMS integrating SOA and Web 2.0 to provide open CMS APIs for three-screen service to IPTV providers or the third party providers. We show that according to normalized costs the proposed user centric CMS for three-screen service is much more cost-effective than the conventional CMS. Therefore, IPTV providers can reduce the depletion of CMS's processing costs and storage as well as network bandwidth for download of oversized content. Social Web TV supporting three-screen service for open IPTV can automatically supply the optimal content using a device-aware function of the social Web TV and content mediation of user centric CMS.

\begin{figure}[t]
\begin{center}
\epsfxsize=6.46cm \leavevmode\epsfbox{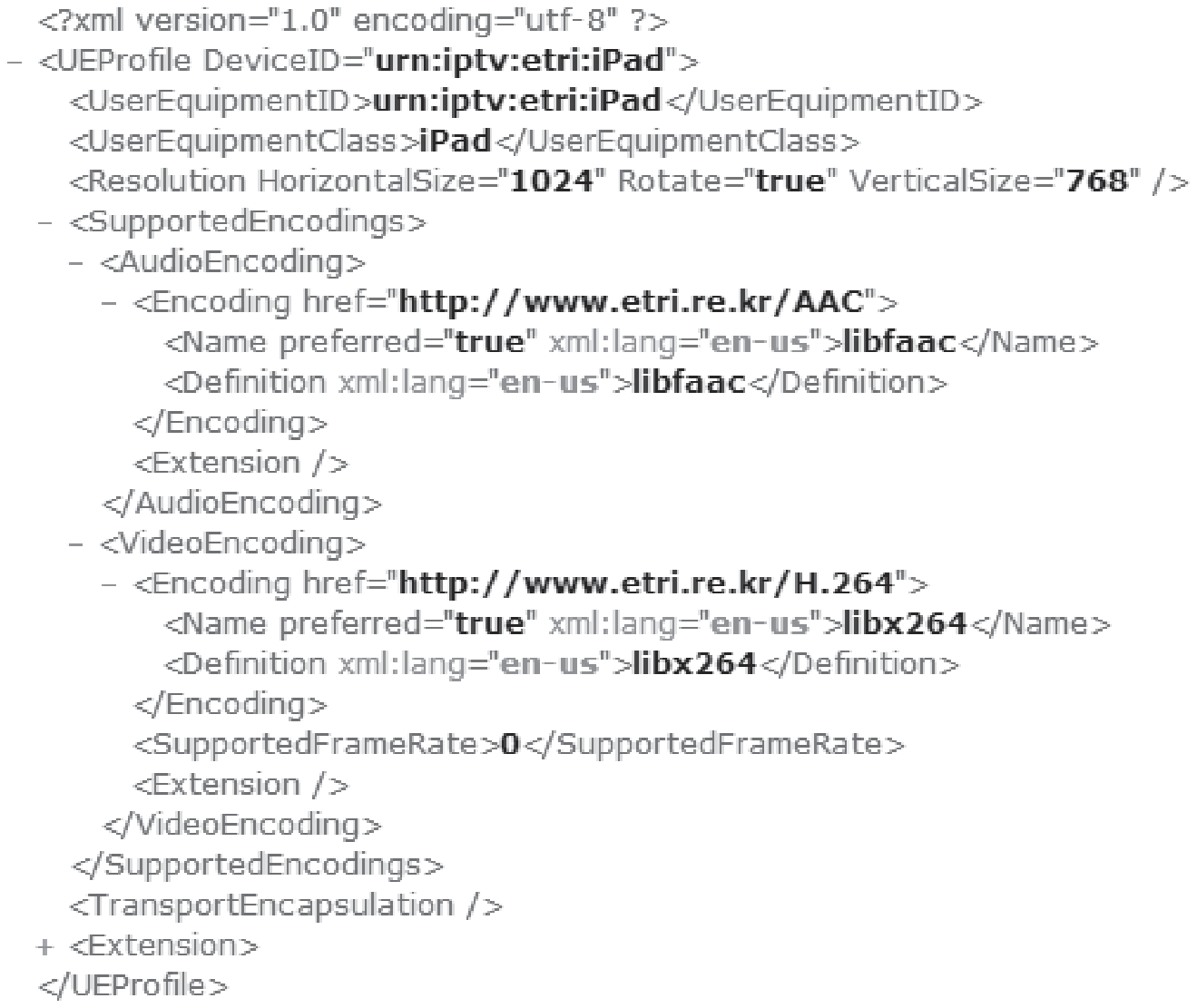}
\caption{An example of device profile for iPad.}
\label{fig:device profile}
\end{center}
\end{figure}

\begin{figure}[t]
\begin{center}
\epsfxsize=6.46cm \leavevmode\epsfbox{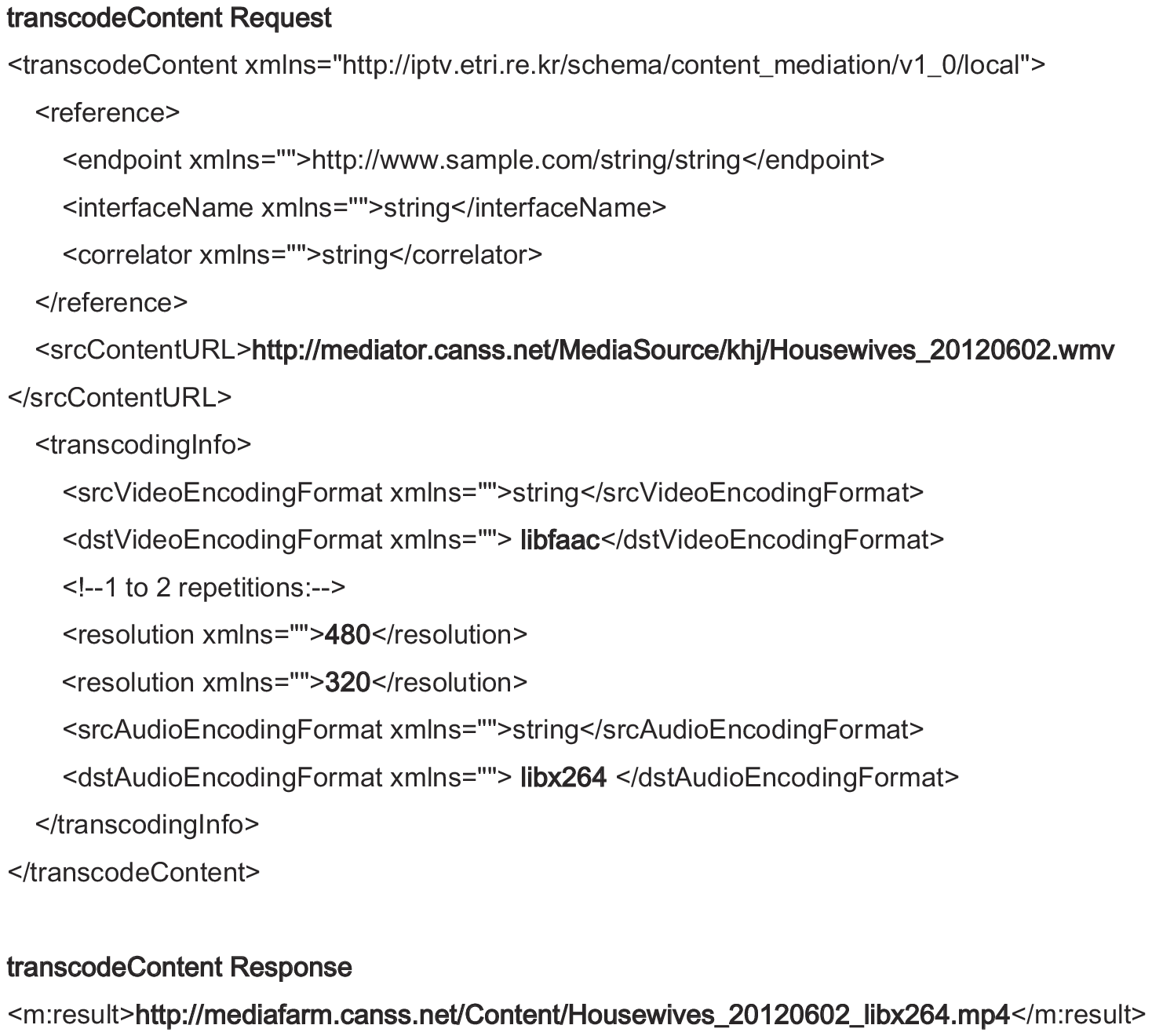}
\caption{SOAP messages of ContentMediation.transcodeContent of Table \ref{tab:core Web services}.}
\label{fig:ws_transcodeContent}
\end{center}
\end{figure}

Furthermore, these Web services of the user centric CMS can be utilized not only for media syndication but also for event-driven applications such as applications using sensors. In this case, the distributed event generators are matched to the content feeder and the central application is matched to social Web TV, which uses data from the event aggregation. We anticipate that this work will be helpful to studies on the next generation open IPTV and Web of Objects (WoO).
\appendix
We describe the interfaces and operations of core Web services for the user centric CMS in Table \ref{tab:core Web services}. An example for device profile of iPad and SOAP messages for ContentMediation.transcodeContent in Table \ref{tab:core Web services} are shown at Fig. \ref {fig:device profile} and Fig. \ref {fig:ws_transcodeContent}, respectively.
\begin{table}[t]
\caption{Interfaces and operations of core Web services for the user centric CMS.}
\scriptsize
\label{tab:core Web services}
\begin{center}
\begin{tabular}{ p{2cm} |  p{3cm} | p{2cm} }
    \hline
    Interfaces & Operations & Input \\ \hline
    ContentAggregation & aggregateContent & reference \\ \cline{3-2} & & feedURL \\ \cline{3-2} & & id \\ \cline{3-2} & & password \\ 
\cline{2-2} \cline{3-2} 
    & getContentAggregationStatus & eventIdentifier \\ \hline
    ContentMediation & transcodeContent & reference \\ \cline{3-2} & & srcContentURL \\ \cline{3-2} & & transcodingInfo \\ 
\cline{2-3} \cline{3-3} 
    & transformMetadata & reference \\ \cline{3-3} & & srcMetadataURL \\ \cline{3-3} & & transformationRule \\ 
\cline{2-3} \cline{3-3}  
    & getContentTranscodingStatus & eventIdentifier \\ 
\cline{2-3} \cline{3-3} 
    & isExistContent & srcContentURL \\ \cline{3-3} & & transcodingInfo \\ \cline{3-3} & & originalContent \\ \hline 
    ContentDeployment & uploadContent & reference \\ \cline{3-3} & & srcLocation \\ \cline{3-3} & & dstLocation \\ 
\cline{2-3} \cline{3-3}
    & getContentUploadingStatus & eventIdentifier \\ \cline{2-3} \cline{3-3} 
    & updateContent & referenxe \\ \cline{3-3} & & srcLocation \\ \cline{3-3} & & dstLocation \\ 
\cline{2-3} \cline{3-3}  
    & getContentUpdatingStatus & eventIdentifier \\ \cline{2-3} \cline{3-3}  
    & deleteContent & reference \\ \cline{3-3} & & crid \\ \hline 
\end{tabular}
\end{center}
\end{table}
\vspace{10pt}

\bibliographystyle{jcn}

\epsfysize=3.2cm
\begin{biography}{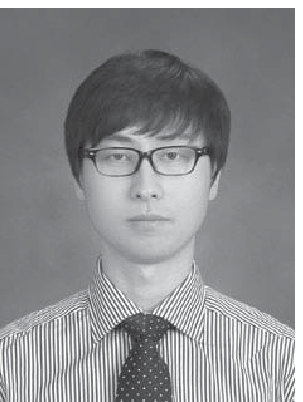}{Seung Hyun Jeon}
received a M.S. degree from Korea Advanced Institute of Science and Technology (KAIST) in 2009 and is currently a Ph.D. student at KAIST. Before joining the doctoral program, he worked as research engineer at the Electronics and Telecommunication Research Institute (ETRI) from early-2010 to mid-2011. He has contributed articles to ITU-T Study Group (SG) 13 Q8, Q9, Q12, Q14, and Q16. Specially, in Q 16, he joined as editor to consent ITU-T Recommendation Y.3022 (measuring energy in networks), which was approved in August 2014. He has studied heterogeneous wireless access networks, multi-connection, and an open IPTV platform in next generation network (NGN). His current research interests include bio-inspired communications and energy efficient solutions for 5G and Smart Grid. 
\end{biography}
\epsfysize=3.2cm
\begin{biography}{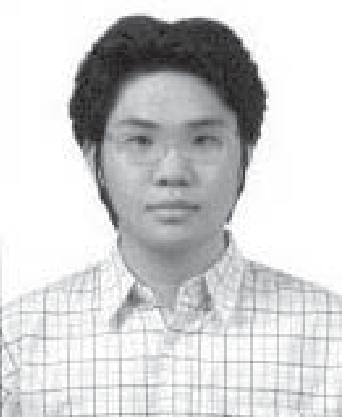}{Sanghong An} received a M.S. degree from KAIST in 2012 and is currently a Ph.D. student at KAIST. He received his bachelor's degree in computer science from KAIST in 2010. He has studied development for IPTV, open IPTV platform, and device Web in next generation network (NGN). His main research interests include Web engineering, Web of Objects, and convergence services in the Internet of Things (IoT), which have primarily been studied in the World Wide Web Consortium (W3C) and Internet Engineering Task Force (IETF).   
\end{biography}
\epsfysize=3.2cm
\begin{biography}{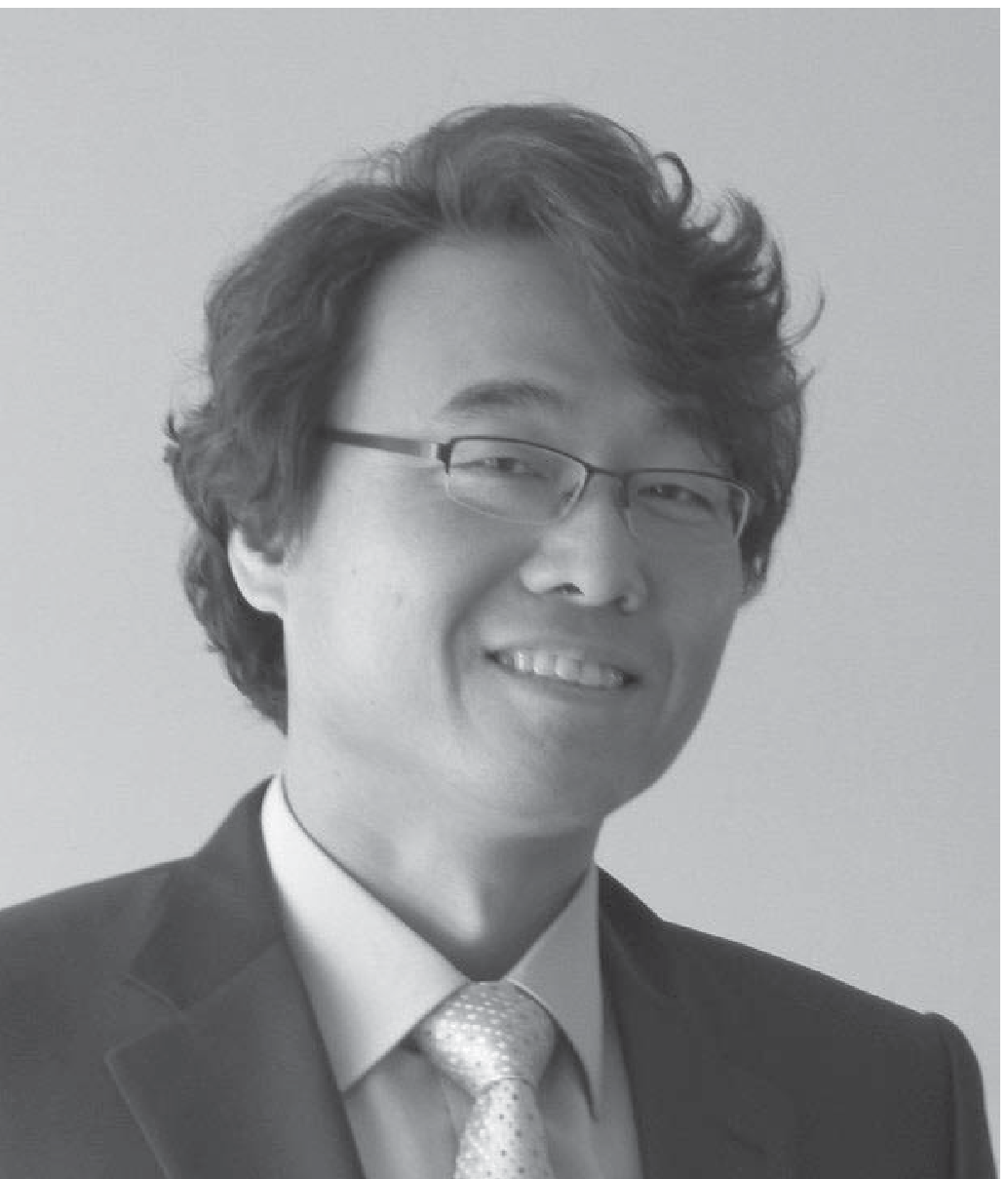}{Changwoo Yoon} received a B.S. degree from Sogang University, Seoul, Korea, in 1990. He received a M.S. degree from POSTECH, Pohang, Korea, in 1992. He received a Ph.D. degree in Computer \& Information Science \& Engineering from University of Florida, US, in 2005. He is currently a principal researcher in the consilience-based research section, future technology research department, ETRI, and an adjunct professor at UST. His current research interests include N-Screen, IPTV, Cloud computing, SOA, service creation/delivery technology, and information retrieval. 
\end{biography}
\epsfysize=3.2cm
\begin{biography}{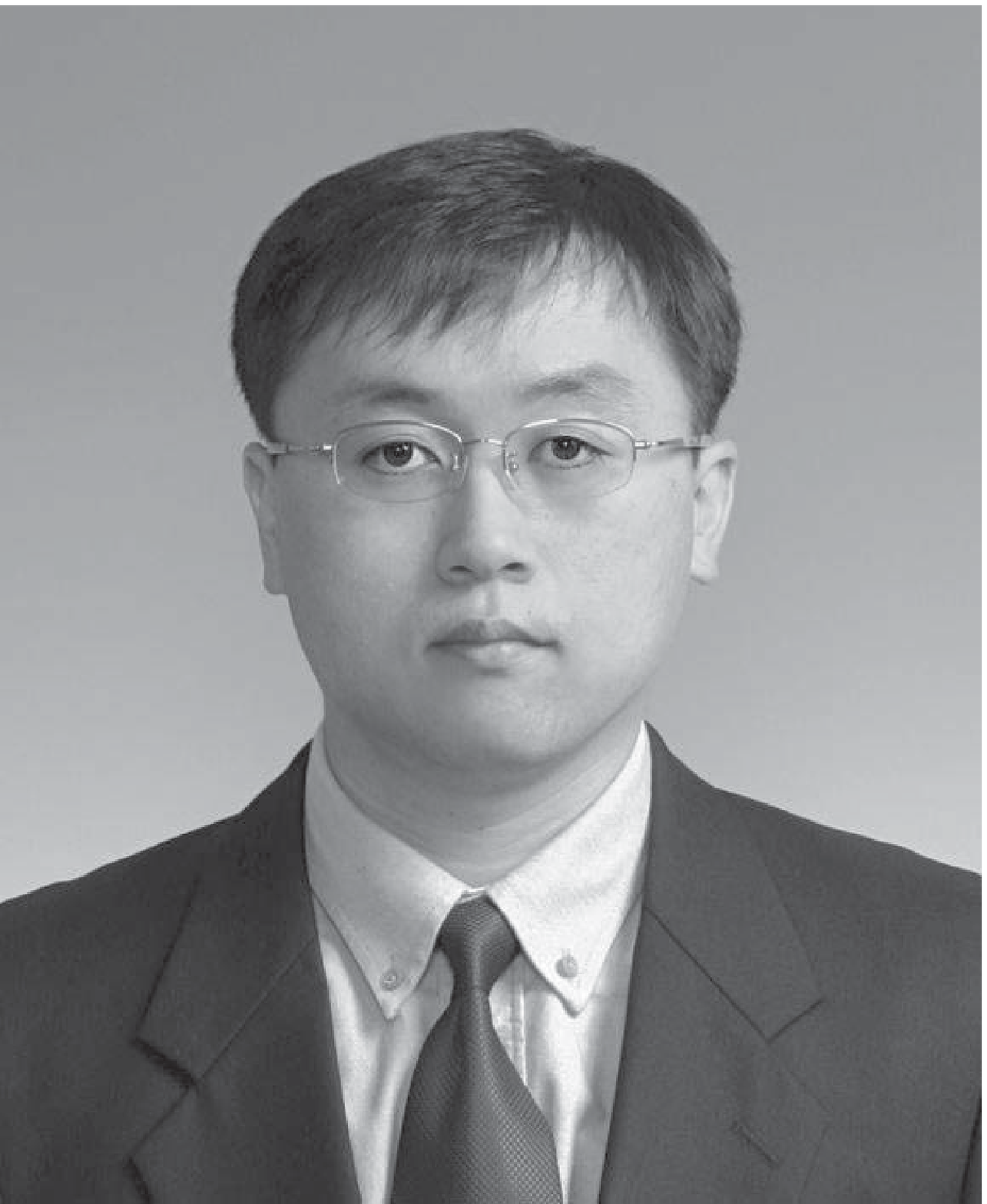}{Hyun-woo Lee} received M.S. and Ph.D. degrees in 1995 and 2005, respectively, from Korea Aerospace University (KAU). He is currently a principal research engineer and section leader in the cloud media networking research section, intelligent convergence technology research department, ETRI. His main research interests include heterogeneous wireless access networks, Mobile P2P, and an open IPTV platform in NGN. His current research interests also include cloud computing and platform.
\epsfysize=3.2cm
\begin{biography}{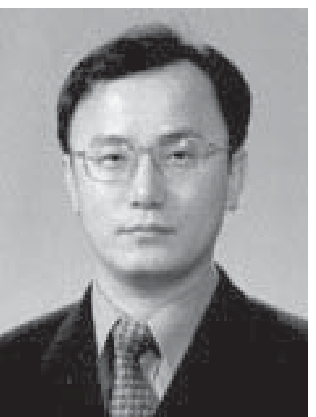}{Junkyun Choi} received M.S. (Eng.) and Ph.D. degrees in 1985 and 1988, respectively, in electronics engineering from KAIST. From June 1986 to December 1997, he was with ETRI. In January 1998 he joined the Information and Communications University (presently, KAIST), Korea as a professor. His research interests are concerned with broadband network architecture and technologies with particular emphasis on performance and protocol problems. Since 1993 he has contributed to consenting ITU Recommendations as a Rapporteur or Editor in ITU-T SG 13 on the ATM, MPLS (Y.1281), NGN (Y.2002, Y.2054), IPTV, and energy saving network (Y.3022, approved in August 2014). He is a Senior Member of IEEE, an executive member of The Institute of Electronics Engineers of Korea (IEEK), an Editor Board Member of Korea Information Processing Society (KIPS), and a Life member of Korea Institute of Communication Science (KICS). His main research interests include Web of objects, open IPTV platform, energy saving networks, measurement platform, and Smart Grid. He has published more than 100 international papers and patent applications in his research areas.  
\end{biography} 
\end{biography}

\end{document}